\documentclass[twocolumn,prl,showpacs,superscriptaddress,amsmath,citeautoscript]{revtex4}
\usepackage{graphicx}
\usepackage{bm}
\usepackage{amssymb}

\begin{document}

\title{Surface spin flip probability of mesoscopic Ag wires}

\author{G.\ Mihajlovi\'{c}}
\email{mihajlovic@anl.gov}
\affiliation{
Materials Science Division, Argonne National Laboratory, Argonne, IL 60439\\}
\author{J.\ E.\ Pearson}
\affiliation{
Materials Science Division, Argonne National Laboratory, Argonne, IL 60439\\}
\author{S.\ D.\ Bader}
\affiliation{
Materials Science Division, Argonne National Laboratory, Argonne, IL 60439\\}
\affiliation{
Center for Nanoscale Materials, Argonne National Laboratory, Argonne, IL 60439\\}
\author{A.\ Hoffmann}
\affiliation{
Materials Science Division, Argonne National Laboratory, Argonne, IL 60439\\}
\affiliation{
Center for Nanoscale Materials, Argonne National Laboratory, Argonne, IL 60439\\}

\date{\today}

\pacs{73.23.-b, 75.40.Gb, 85.75.-d}

\begin{abstract}
Spin relaxation in mesoscopic Ag wires in the diffusive transport regime is studied via nonlocal spin valve and Hanle effect measurements performed on permalloy/Ag lateral spin valves. The ratio between momentum and spin relaxation times is not constant at low temperatures.  This can be explained with the Elliott-Yafet spin relaxation mechanism by considering the momentum surface relaxation time as being temperature dependent. We present a model to separately determine spin flip probabilities for phonon, impurity and surface scattering and find that the spin flip probability is highest for surface scattering.
\end{abstract}

\maketitle

Understanding how confinement influences physical properties is crucial for advancing nanotechnology \cite{Bader2006}. Numerous studies have shown that when one or more dimensions of a structure become comparable to a characteristic length scale of a physical process in question ({\em e.g.}, a mean free path for electron transport) even classical boundary or surface effects can give rise to dramatically different behavior than that expected for the same bulk material. Examples include magnetoresistance in semiconductor nanostructures (negative $vs.$ positive in the bulk) \cite{Thornton1989} or thermal conductivities in Si nanowires (orders of magnitude reduction compared to bulk Si) \cite{Boukai2008}.  In contrast, confinement effects are less evident in metallic transport due to inherently short mean free paths but often manifest themselves in optical properties \cite{Barnes2003}. An important question to be addressed in  $spintronics$ \cite{Zutic2004} is how does the size of a spin conductor or the surface conditions affect the transport of spin currents?  Due to the relatively long spin diffusion length compared to the mean free path, confinement effects can be more pronounced in spin transport, even in metallic structures.  So far, experiments performed with metallic lateral spin valve (LSV) structures \cite{Jedema2001, Ji2004}, where pure spin currents in a non-magnetic normal metal (N) are generated by diffusion of the non-equilibrium spin accumulation injected from a ferromagnet (F)\cite{Johnson-Silsbee1985}, have focused mostly on determining spin diffusion lengths $l_s$ and spin injection efficiencies for various combinations of F/N materials, without quantifying contributions of different scattering mechanisms to the spin relaxation. In particular, to what extent does confinement affect the spin relaxation time $\tau_s$ \cite{Jedema2003}? In this Letter we present a model, based on the Elliott-Yafet (EY) mechanism of spin relaxation \cite{Elliott1954, Yafet1963}, to separately quantify spin flip probabilities for phonon, impurity and surface scattering in mesoscopic metal wires in the diffusive transport regime. By studying spin transport in permalloy (Py)/Ag LSVs we find that the spin flip probability is highest for electron scattering from the Ag surface. Our model can also explain recent experimental results on temperature $T$ \cite{Kimura2008} as well as thickness dependence of $l_s$ in mesoscopic Cu wires \cite{Erekhinsky2009}.

\begin{figure}
\includegraphics[scale=0.48, bb=18 4 487 502]{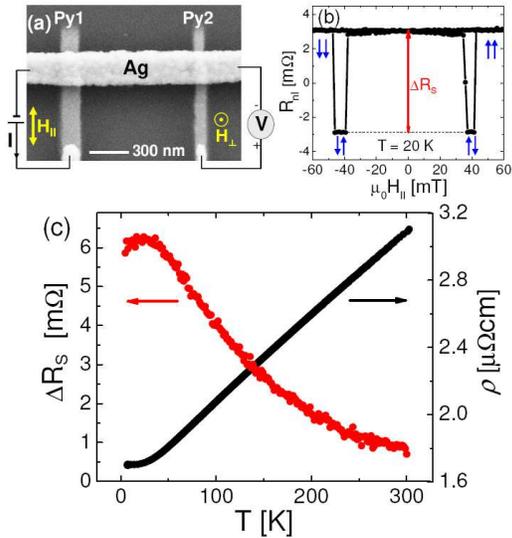}
\caption{(Color online) (a) An SEM image of a Py/Ag LSV device adapted to show the nonlocal measurement configuration. Also shown are the directions of $H_\parallel$ and $H_\perp$ applied in NLSV and Hanle effect measurements, respectively. (b) $R_{nl}$ {\em vs.}\  $H_\parallel$ at 20~K. Corresponding $M$ orientations of the Py electrodes are shown as blue arrows, while the total $\Delta R_s$ signal is highlighted in red. (c) $T$ dependencies of $\Delta R_s$ and $\rho$.}
\label{fig:nonlocal}
\end{figure}

The Py/Ag LSV devices were fabricated on a SiN (100~nm)/Si substrate by e-beam lithography and shadow mask e-beam evaporation. A scanning electron microscope (SEM) image of a central region of the device is shown in Fig.~1(a). The two Py electrodes Py1 and Py2 were both 25~nm thick and had widths of 130 and 80~nm respectively, while the bridging Ag wire was 260~nm wide and $d = 80$~nm thick. The center to center distance $L$ between Py electrodes was 705~nm. Nonlocal spin valve (NLSV) and Hanle effect measurements were performed by applying a {\em dc} current $I$ = $\pm$~0.3~mA  from Py1 into the left part of the Ag wire and measuring the voltage $V$ between Py2 and the right end of the Ag wire as a function of parallel $H_\parallel$ and perpendicular $H_\bot$ magnetic fields, respectively [see Fig.~1(a)].

Figure 1(b) shows the result of the NLSV measurement at 20~K. The dips in the nonlocal resistance, $R_{nl} = V/I$, due to spin accumulation in Ag, upon switching the magnetization $M$ orientation of Py1, are clearly observed.  The magnitude of the difference between the $R_{nl}$ values measured for parallel and antiparallel $M$ orientation of Py electrodes, $\Delta R_s$, is $\sim 6.1$~m$\Omega$, which is a large signal  given that $L$ = 705~nm \cite{Godfrey-Johnson2006, Kimura2007}. Also note that the values of $R_{nl}$ for parallel and antiparallel  $M$ orientation of the Py electrodes are almost perfectly symmetric with respect to zero, meaning that they are  due to pure spin transport without parasitic ohmic signals \cite{Casanova2009, Mihajlovic2009}.

The large $\Delta R_s$ values facilitate measurements of its $T$ dependence  by measuring $R_{nl}(T)$  for parallel and antiparallel remanent $M$ orientations of the Py electrodes, respectively. Figure~1(c) shows the $T$ dependence of $\Delta R_s = R_{nl}^{\uparrow\uparrow}-R_{nl}^{\downarrow\uparrow}$ and the corresponding conventional electrical resistivity $\rho$ of the Ag wire. $\Delta R_s$ is non-monotonic at low $T$ despite the monotonic decrease of $\rho$ [see Fig.~1(c)]. This behavior is consistently observed in all measured samples. In addition, similar behavior has also been observed in the case of Py/Cu LSVs \cite{Kimura2008} and was attributed to the reduction of $l_s$ in Cu due to surface scattering.

In order to determine whether a similar  physical mechanism also causes the behavior observed in our sample, we turned to Hanle effect measurements, since they avoid variabilities between different samples, which are unavoidable in thickness-dependent studies. For Hanle measurements a combined effect of spin precession, relaxation and dephasing leads to a characteristic dependence of $\Delta R_s$ on $H_\bot$, the Hanle resistance $R_H$, given as: \cite{Jedema2002}
\begin{equation}
   R_H^{\uparrow\uparrow}(H_\perp) = \frac{P^2 \rho D}{A}\int_0^\infty \mathcal{P}(t) \cos(\omega_L t) \exp \left(-\frac{t}{\tau_s}\right)dt,
  \label{eq:2}
\end{equation}
for parallel $M$ orientation of Py electrodes, and $R_H^{\downarrow\uparrow}(H_\perp) = - R_H^{\uparrow\uparrow}(H_\perp)$ for antiparallel one. Here, $\omega_L = g \mu_B \mu_0 H_\perp / \hbar$ is the Larmor frequency ($g$ is the Lande factor, $\mu_B$ is the Bohr magneton, $\mu_0$ is the magnetic permeability of free space) and $\mathcal{P}(t) = (1/\sqrt{4\pi D t})\exp(-L^2/4Dt)$ is the probability distribution of traveling times $t$ of the injected spins from Py1 to Py2. Thus the Hanle effect measurements  are used to determine separately the injected spin polarization $P$ and $\tau_s$ by fitting  $R_H$~$vs.$~$H_\perp$ data to Eq.~(1), if one knows the diffusion constant $D$, $\rho$, $L$, and the cross sectional area of the wire, $A$.

\begin{figure}
\includegraphics[scale=0.62, bb=0 0 384 445]{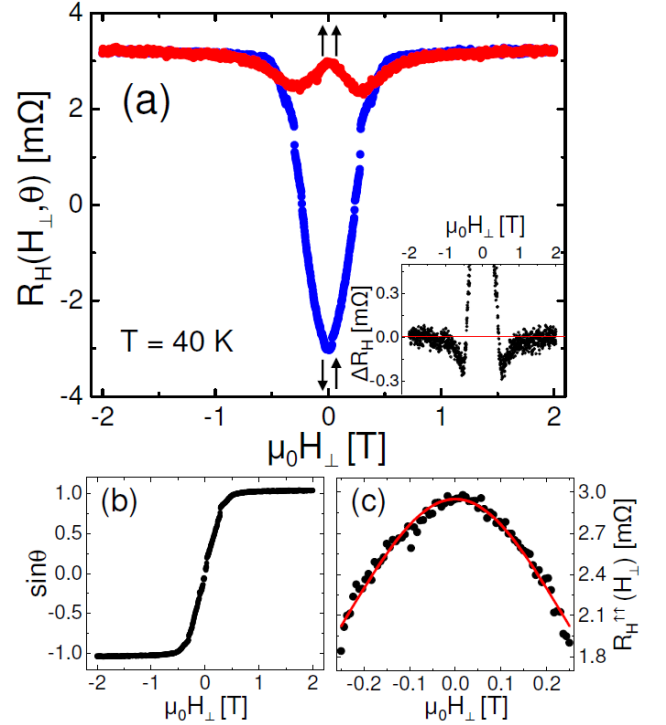}
\caption{(Color online) (a) Hanle signals measured at 40~K for parallel ($\uparrow\uparrow$, red) and antiparallel ($\downarrow\uparrow$, blue) $M$ orientations of Py electrodes. Inset: $\Delta R_H = R_H ^{\uparrow\uparrow}-R_H ^{\uparrow\downarrow}$. Red line marks $\Delta R_H =0$ for clarity. (b) $\sin\theta$ {\em vs.}\ $\mu_0 H_\perp$ obtained from data shown in (a) by using Eq.~(3). (c) Hanle signal (symbols) at 40~K from the data shown in (a) by using Eq.~(4). Best fit according to Eq.~(1) shown as a red line.}
\label{fig:nonlocal}
\end{figure}

Figure~2(a) shows representative $R_H$ data, obtained at 40~K for Py electrodes prepared in parallel (red) and antiparallel (blue) $M$ configurations. The difference between the two signals (not shown) exhibits an oscillating sign change as a function of $H_\perp$, as expected from the spin precession. However, the behavior is more complicated than expected from Eq.~(1). The striking feature of the data in Fig.~2(a) is the asymmetric shape of the two curves with respect to $R_{H}$~=~0, which has not been observed in previously reported Hanle effect measurements in Al \cite{Jedema2002, Valenzuela2006}. To understand this asymmetry, observed at all $T$s, one has to take into account that in addition to precession and dephasing of the spin accumulation, the measured signal also depends on the orientation of the $M$ of the Py electrodes with respect to the substrate plane. Namely, $M$ inevitably tilts in the perpendicular direction due to application of $H_\perp$. This decreases the fraction of precessing spin accumulation, and tends to restore the $R_H$ signal to its initial value of $\Delta R_s/2$ for parallel $M$ orientations.   When  this effect is taken into account,  $R_H$ can be expressed as: \cite{Jedema2002}
\begin{equation}
  R_H^{\uparrow\uparrow (\downarrow\uparrow)}(H_\bot,\theta)= \pm R_H^{\uparrow\uparrow}(H_\bot) \cos^2(\theta)+|R_H(0)| \sin^2(\theta),
  \label{eq:3}
\end{equation}
with $"+"$ and $"-"$ signs corresponding to the $\uparrow\uparrow$ and $\downarrow\uparrow$ case, respectively. Here $\theta$ is the angle between the substrate plane and the direction of  $M$. Based on Eq.(2):
\begin{equation}
  \sin^2(\theta) = \frac{R_H^{\uparrow\uparrow}(H_\bot,\theta)+R_H^{\uparrow\downarrow}(H_\bot,\theta)}{2|R_H(0)|}.
  \label{eq:5}
\end{equation}

Figure~2(b) shows the dependence of $\sin \theta$ on $\mu_0 H_\bot$ obtained using Eq.~(3) for the case shown in Fig.~2(a). For sufficiently low values of $\mu_0 H_\bot$ the dependence is linear, but saturates above $\sim \pm 0.5$~T. This dependence is consistent with the Stoner-Wohlfarth model \cite{Stoner1948} for coherent $M$ rotation with fields applied along a hard-axis direction. The slope of 2.7~T$^{-1}$ for $\sin \theta$ around zero field corresponds to a demagnetizing factor $N = 0.37$, taking $\mu_0 M_s$ = 1~T for Py.  This value agrees reasonably well with the literature one of $N = 0.5$ taking into account that the latter is defined for an infinitely long wire. Therefore, we conclude that the asymmetry in the measured $R_H$ curves  arises from the tilting of the Py magnetizations.

From the data, and Eq.~(2), we can extract the Hanle signal:
\begin{equation}
  R_H^{\uparrow\uparrow}(H_\bot)= |R_H(0)| \frac{R_H^{\uparrow\uparrow}(H_\bot,\theta)-R_H^{\downarrow\uparrow}(H_\bot,\theta)}{2|R_H(0)|-R_H^{\uparrow\uparrow}(H_\bot,\theta)-R_H^{\downarrow\uparrow}(H_\bot,\theta)},
  \label{eq:6}
\end{equation}

while $R_H^{\downarrow\uparrow}(H_\bot) = -R_H^{\uparrow\uparrow}(H_\bot)$. Figure~2(c) shows $R_H^{\uparrow\uparrow}(H_\bot)$ obtained using Eq.~(4) for 40~K. The best fit according to Eq. (1) is shown as a red line. Only the region between $\pm$~0.25~T is used since $R_H^{\uparrow\uparrow}(H_\bot)$ is not well defined for higher fields due to the denominator in Eq.~(4) being close to zero. The fit gives $P$ = 0.207 and  $\tau_s$ = 14.4 ps, corresponding to $l_s$~=~564~nm. These values agree well with the ones reported in Ref.~\cite{Godfrey-Johnson2006} (note that difference in $l_s$ scales with the difference in $\rho$), but are considerably different than those reported in Ref.~\cite{Kimura2007}. This suggests that determination of $l_s$ based on  the transparent interface model used in Ref.~\cite{Kimura2007} is not appropriate, due to the presence of an insulating oxide layer at the Py/Ag interface \cite{Mihajlovic2010}. We repeated the above procedure to determine $P$ and $\tau_s$ in the range from 4.5 to 200~K, where the Hanle effect was observed.

\begin{figure}
\includegraphics[scale=0.45, bb=0 0 545 325]{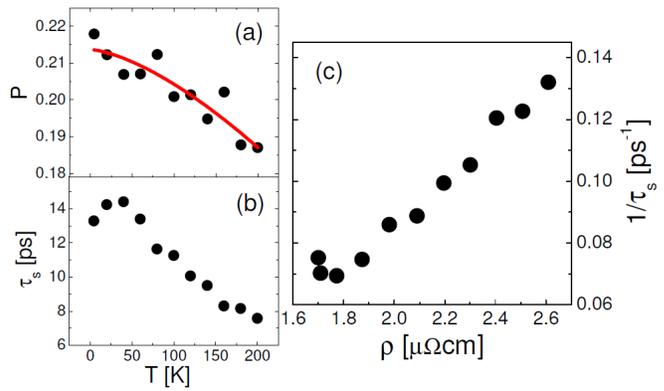}
\caption{(a) $T$ dependence of $P$. The red line is fit to the Bloch model of thermally excited spin waves (see text), with $P_0 = 0.214 \pm 0.003$ and $\eta = (4.0 \pm 0.6)\times 10^{-5}$~K$^{-3/2}$. (b) $T$ dependence of $\tau_s$. (c) Dependence of spin relaxation rate on resistivity of the Ag wire.}
\label{fig:nonlocal}
\end{figure}

Figure~3(a) shows the $T$ dependence of $P$. This dependence is monotonic and can be fitted to the Bloch model of thermally excited spin waves, $i.e.$ $P(T)=P_0(1-\eta T^{3/2}$ as expected \cite{Shang1998}. In contrast,  $\tau_s$ exhibits a maximum around $\sim$40~K  [see Fig.~3(b)] and then slightly decreases with decreasing  $T$. This confirms that the observed non-monotonic $T$ dependence of $\Delta R_s$ [see Fig.~1(c)] is due to the reduction of $\tau_s$, and hence $l_s$, at low $T$. Similar behavior has been observed for Cu \cite{Kimura2008}. The dependence of spin relaxation rate $1/\tau_s$ on $\rho$ is shown in Fig. 3(c). $1/\tau_s$ increases linearly with $\rho$ above $\sim$40~K, as expected from the EY mechanism of spin relaxation \cite{Elliott1954, Yafet1963, Fabian-DasSarma1999}, but it exhibits a minimum around this $T$, and then slightly increases with decreasing $\rho$. We point out that the surface spin relaxation \cite{Lindelof1986, Wang1995} based on the  Fuchs model \cite{Sondheimer1952} with a $T$-independent surface momentum relaxation time  $\tau_e^S$ cannot \emph{quantitatively} explain the nonlinear dependence of $1/\tau_s$ on $\rho$, let alone the upturn of $1/\tau_s$ at low $T$. The  discrepancy can be resolved by invoking the concept of a $T$-dependent  $\tau_e^S$ \cite{Watts1968, Wang1973, Zamaleev1978}. Namely, when transport in the wire is diffusive, the $\tau_e^S$, which is the average time it takes an electron to diffuse over the distance $d$  is given as  $\tau_e^S = \gamma d^2/D_B$. Here, $\gamma$ is the averaging coefficient  and $D_B = (1/3) v_F^2 \tau_e^B$ is the $T$-dependent bulk diffusion constant, with $v_F = 1.39 \times 10^6$~m/s  and $\tau_e^B$ being the electron Fermi velocity and the bulk momentum relaxation time, respectively, . The latter is determined by scattering within the bulk of the wire and can be defined as $(\tau_e^B)^{-1} = (\tau_e^{ph})^{-1} + (\tau_e^{imp})^{-1}$, where $\tau_e^{ph}$ and $\tau_e^{imp}$ are momentum relaxation times for electron scattering from phonons and impurities (including the grain boundaries), respectively.  Thus, the total momentum relaxation time $\tau_e$:
\begin{equation}
  \frac{1}{\tau_e} = \frac{1}{\tau_e^{ph}}+ \frac{1}{\tau_e^{imp}} + \frac{1}{3\gamma}\left(\frac{v_F}{d}\right)^2 \tau_e^B.
  \label{eq:6}
\end{equation}
Associating  to each scattering process its corresponding spin flip probability, {\em i.e.}, $\epsilon_{ph}$, $\epsilon_{imp}$ and $\epsilon_{S}$, and following the EY proportionality between momentum and spin relaxation times \cite{Fabian-DasSarma1999}, we find
\begin{eqnarray}
  \frac{1}{\tau_s}& = &\frac{\epsilon_{ph}}{\tau_e^{ph}}+ \frac{\epsilon_{imp}}{\tau_e^{imp}} + \frac{1}{3\gamma}\left(\frac{v_F}{d}\right)^2 \epsilon_{S}\tau_e^B \\
  &=&\frac{\epsilon_{ph}}{\tau_e^B} + \frac{1}{3\gamma}\left(\frac{v_F}{d}\right)^2 \epsilon_{S}\tau_e^B + \frac{\epsilon_{imp}-\epsilon_{ph}}{\tau_e^{imp}}.
  \label{eq:6}
\end{eqnarray}
Equation ~(7) naturally explains the nonlinearity in $1/\tau_s$ {\em vs.}\ $\rho$ (since $\tau_e^B$ within this model depends nonlinearly on $\rho$) and, even more significantly, it describes the upturn in $1/\tau_s$ at low $T$, since the first and second terms in Eq.~(7) have different dependencies on $\rho$.   Also, based on  Eq.~(7), one can determine $\epsilon_{ph}$, $\epsilon_{imp}$ and $\epsilon_{S}$ by  fitting the  $1/\tau_s$ $vs.$ $1/\tau_e^B$ dependence, which can be obtained by performing $T$-dependent spin transport measurements,  such as is shown in Fig.~3.

\begin{figure}
\includegraphics[scale=0.73, bb=17 10 280 225]{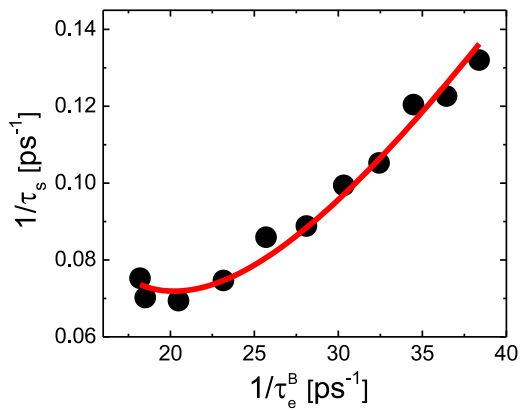}
\caption{(Color online) Dependence of spin relaxation rate in the Ag wire on the bulk momentum relaxation rate (symbols). The fit to the data according to Eq.~(6) is shown as a red line.}
\label{fig:nonlocal}
\end{figure}

Figure~4 shows the plot of $1/\tau_s$ $vs.$ $1/\tau_e^B$ (black circles), where $\tau_e^B$ is obtained using Eq.~(5) as $\tau_e^B = (1-\sqrt{1-4a\tau_e^2})/(2a\tau_e)$, with  $a =  (v_F/d)^2/3\gamma$ = 179.3~ps$^{-2}$ was obtained by determining $\gamma = 0.563$ using Dingle's effective mean free path model for completely diffuse electron surface scattering \cite{Dingle1950}, and $\tau_e = m/(ne^2\rho)$ was determined from the measured $\rho$ using $n = 5.85 \times 10^{28}$~m$^{-3}$. The best fit according to Eq.~(7) is shown as a red line. We find $\epsilon_{ph} = (7.5 \pm 1.3) \times 10^{-3}$, $\epsilon_S = (1.7 \pm 0.4) \times 10^{-2}$ and $\epsilon_{imp} = (-5.0 \pm 5.2) \times 10^{-3}$. These values show that the spin flip probability is highest for  surface scattering and weakest for scattering from impurities ($< 2\times 10^{-4}$)\endnote{Note that the negative sign of the fitted $\epsilon_{imp}$ is meaningless and within the error bar.  Thus, we do not have sufficient sensitivity to determine $\epsilon_{imp}$.}. Also note that the value for $\epsilon_{ph}$ agrees well with a previously reported value of $\epsilon_{ph} = 2 \times 10^{-3}$ for Cu wires ($\sim 4\times$ smaller value, consistent with Cu being a lighter element) \cite{Jedema2003, Beuneu1976} as well as the value of $\epsilon_{ph} = 2.9 \times 10^{-3}$ obtained from electron spin resonance on bulk Ag \cite{Beuneu1976}.

Based on this model, we can also predict the thickness dependence of  $l_s$. Multiplying Eq.~(7) by $1/D$ one finds
\begin{equation}
  l_s = \frac{d}{\sqrt{\alpha d^2 + \beta}},
  \label{eq:6}
\end{equation}
where in the low $T$ limit, when $\tau_e^B \simeq \tau_e^{imp}$,  $\alpha = \epsilon_{imp}/(D \tau_e^{imp})$ and $\beta = (v_F^2 \epsilon_S \tau_e^{imp})/(3\gamma D)$. In the high $T$ limit, where phonon scattering is non-negligible, $\alpha = \epsilon_{imp}/(D \tau_e^{imp}) + \epsilon_{ph}/(D \tau_e^{ph})$, and $\beta$ is the same. These relations in principle can be used to determine $\epsilon_{ph}$, $\epsilon_{imp}$ and $\epsilon_{S}$ by fitting the thickness dependence of the spin diffusion length obtained for low and high $T$.  However, such measurements require several samples, which introduce additional experimental uncertainties.

In conclusion, we have studied the spin relaxation in a mesoscopic Ag wire in the diffusive transport regime, and observed a nonlinear dependence of the spin relaxation rate on resistivity. This observation cannot be explained quantitatively with the Elliott-Yafet mechanism of spin relaxation by a conventional approach, which considers surface spin relaxation as being temperature independent.  We present a model that explains these observations with the Elliott-Yafet mechanism by adding a temperature dependence to the surface relaxation.  This enables us to quantify spin flip probabilities for phonon, impurity and surface scattering respectively. We find that the spin flip probability of the Ag wire is strongest for electron scattering from surfaces.

We thank R. Winkler, K. Vyborny, and O. Mosendz for stimulating discussions, and  L. Ocola and R. Divan for assistance with nanofabrication. This work was supported by the U.S.\ Department of Energy, Office of Science, Basic Energy Sciences, under contract No. DE-AC02-06CH11357.


\begin{thebibliography}{35}
\expandafter\ifx\csname natexlab\endcsname\relax\def\natexlab#1{#1}\fi
\expandafter\ifx\csname bibnamefont\endcsname\relax
  \def\bibnamefont#1{#1}\fi
\expandafter\ifx\csname bibfnamefont\endcsname\relax
  \def\bibfnamefont#1{#1}\fi
\expandafter\ifx\csname citenamefont\endcsname\relax
  \def\citenamefont#1{#1}\fi
\expandafter\ifx\csname url\endcsname\relax
  \def\url#1{\texttt{#1}}\fi
\expandafter\ifx\csname urlprefix\endcsname\relax\def\urlprefix{URL }\fi
\providecommand{\bibinfo}[2]{#2}
\providecommand{\eprint}[2][]{\url{#2}}

\bibitem[{\citenamefont{Bader}(2006)\citenamefont{S. D. Bader}}]{Bader2006}
\bibinfo{author}{\bibfnamefont{S.~D.}~\bibnamefont{Bader}},
  \bibinfo{journal}{Rev.\ Mod.\ Phys.} \textbf{\bibinfo{volume}{78}},
  \bibinfo{pages}{1} (\bibinfo{year}{2006}).

\bibitem[{\citenamefont{Thornton et al.}(1989)\citenamefont{T. J. Thornton, M. L. Roukes, A. Scherer, and B. P. Van de Gaag}}]{Thornton1989}
\bibinfo{author}{\bibfnamefont{T.~J.}~\bibnamefont{Thornton}},
\bibinfo{author}{\bibfnamefont{M.~L.}~\bibnamefont{Roukes}},
\bibinfo{author}{\bibfnamefont{A.}~\bibnamefont{Scherer}},
\bibnamefont{and}
  \bibinfo{author}{\bibfnamefont{B.~P.} \bibnamefont{Van de Gaag}},
  \bibinfo{journal}{Phys.\ Rev.\ Lett.} \textbf{\bibinfo{volume}{63}},
  \bibinfo{pages}{2128} (\bibinfo{year}{1989}).

\bibitem[{\citenamefont{Boukai et al.}(2008)\citenamefont{Akram I. Boukai1,2, Yuri Bunimovich1,2, Jamil Tahir-Kheli1, Jen-Kan Yu1, William A. Goddard III1 & James R. Heath1}}]{Boukai2008}
\bibinfo{author}{\bibfnamefont{A.~I.}~\bibnamefont{Boukai}},
\bibinfo{author}{\bibfnamefont{Y.}~\bibnamefont{Bunimovich}},
\bibinfo{author}{\bibfnamefont{J.}~\bibnamefont{Tahir-Kheli}},
\bibinfo{author}{\bibfnamefont{J.~-K.}~\bibnamefont{Yu}},
\bibinfo{author}{\bibfnamefont{W.~A.}~\bibnamefont{Goddard III}},
\bibnamefont{and}
  \bibinfo{author}{\bibfnamefont{J.~R.} \bibnamefont{Heath}},
  \bibinfo{journal}{Nature\ (London)} \textbf{\bibinfo{volume}{451}},
  \bibinfo{pages}{168} (\bibinfo{year}{2008}).

\bibitem[{\citenamefont{Barnes et~al.}(2003)\citenamefont{Barnes, Dereux, and Ebbesen}}]{Barnes2003}
\bibinfo{author}{\bibfnamefont{W.~L.} \bibnamefont{Barnes}},
  \bibinfo{author}{\bibfnamefont{A.} \bibnamefont{Dereux}},
  \bibnamefont{and}
  \bibinfo{author}{\bibfnamefont{T.~W.} \bibnamefont{Ebbesen}},
  \bibinfo{journal}{Nature (London)} \textbf{\bibinfo{volume}{424}},
  \bibinfo{pages}{824} (\bibinfo{year}{2003}).

\bibitem[{\citenamefont{Zutic et~al.}(2004)\citenamefont{Zutic, Fabian, and Das Sarma}}]{Zutic2004}
\bibinfo{author}{\bibfnamefont{I.} \bibnamefont{\u{Z}uti\'{c}}},
  \bibinfo{author}{\bibfnamefont{J.} \bibnamefont{Hoffmann}},
  \bibnamefont{and}
  \bibinfo{author}{\bibfnamefont{S.} \bibnamefont{Das Sarma}},
  \bibinfo{journal}{Rev.\ Mod.\ Phys.} \textbf{\bibinfo{volume}{76}},
  \bibinfo{pages}{323} (\bibinfo{year}{2004}).

\bibitem[{\citenamefont{Jedema et~al.}(2001)\citenamefont{Jedema, Filip, and van Wees}}]{Jedema2001}
\bibinfo{author}{\bibfnamefont{F.~J.}~\bibnamefont{Jedema}},
\bibinfo{author}{\bibfnamefont{A.~T.}~\bibnamefont{Filip}},
\bibnamefont{and}
  \bibinfo{author}{\bibfnamefont{B.~J.} \bibnamefont{van Wees}},
  \bibinfo{journal}{Nature\ (London)} \textbf{\bibinfo{volume}{410}},
  \bibinfo{pages}{345} (\bibinfo{year}{2001}).

\bibitem[{\citenamefont{Ji et~al.}(2004)\citenamefont{Ji, Hoffmann, Jiang, and Bader}}]{Ji2004}
\bibinfo{author}{\bibfnamefont{Y.} \bibnamefont{Ji}},
  \bibinfo{author}{\bibfnamefont{A.} \bibnamefont{Hoffmann}},
  \bibinfo{author}{\bibfnamefont{J.~S.} \bibnamefont{Jiang}},
  \bibnamefont{and}
  \bibinfo{author}{\bibfnamefont{S.~D.} \bibnamefont{Bader}},
  \bibinfo{journal}{Appl.\ Phys.\ Lett.} \textbf{\bibinfo{volume}{85}},
  \bibinfo{pages}{6218} (\bibinfo{year}{2004}).

  \bibitem[{\citenamefont{Johnson and Silsbee}(1985)\citenamefont{Johnson and Silsbee}}]{Johnson-Silsbee1985}
\bibinfo{author}{\bibfnamefont{M.}~\bibnamefont{Johnson}}
  \bibnamefont{and}
  \bibinfo{author}{\bibfnamefont{R.~H.} \bibnamefont{Silsbee}},
  \bibinfo{journal}{Phys.\ Rev.\ Lett.} \textbf{\bibinfo{volume}{55}},
  \bibinfo{pages}{1790} (\bibinfo{year}{1985}).

\bibitem[{\citenamefont{Jedema et~al.}(2003)\citenamefont{Jedema, Nijboer, Filip, and van Wees}}]{Jedema2003}
\bibinfo{author}{\bibfnamefont{F.~J.}~\bibnamefont{Jedema}},
\bibinfo{author}{\bibfnamefont{M.~S.}~\bibnamefont{Nijboer}},
\bibinfo{author}{\bibfnamefont{A.~T.}~\bibnamefont{Filip}},
\bibnamefont{and}
  \bibinfo{author}{\bibfnamefont{B.~J.} \bibnamefont{van Wees}},
  \bibinfo{journal}{Phys.\ Rev.\ B} \textbf{\bibinfo{volume}{67}},
  \bibinfo{pages}{085319} (\bibinfo{year}{2003}).

\bibitem{Elliott1954}
\bibinfo{author}{\bibfnamefont{R.~J.}~\bibnamefont{Elliott}},
  \bibinfo{journal}{Phys.~Rev.} \textbf{\bibinfo{volume}{96}},
  \bibinfo{pages}{266} (\bibinfo{year}{1954}).

\bibitem[{\citenamefont{Yafet}(1963)}]{Yafet1963}
\bibinfo{author}{\bibfnamefont{Y.} \bibnamefont{Yafet}},~in
\emph{Solid State Physics}, \bibinfo{journal}{edited by F. Seitz and D. Turnbull}
  (\bibinfo{year}{Academic, New York, 1963}),~Vol.~14.

\bibitem[{\citenamefont{Kimura et~al.}(2008)\citenamefont{Kimura, Sato, and Otani}}]{Kimura2008}
\bibinfo{author}{\bibfnamefont{T.}~\bibnamefont{Kimura}},
\bibinfo{author}{\bibfnamefont{T.}~\bibnamefont{Sato}},
  \bibnamefont{and}
  \bibinfo{author}{\bibfnamefont{Y.} \bibnamefont{Otani}},
  \bibinfo{journal}{Phys.\ Rev.\ Lett.} \textbf{\bibinfo{volume}{100}},
  \bibinfo{pages}{066602} (\bibinfo{year}{2008}).

\bibitem[{\citenamefont{Erekhinsky et~al.}(2009)\citenamefont{Mikhail Erekhinsky, Amos Sharoni, Felix Casanova, Ivan K. Schuller}}]{Erekhinsky2009}
\bibinfo{author}{\bibfnamefont{M.}~\bibnamefont{Erekhinsky}},
\bibinfo{author}{\bibfnamefont{A.}~\bibnamefont{Sharoni}},
\bibinfo{author}{\bibfnamefont{F.}~\bibnamefont{Casanova}},
\bibnamefont{and}
  \bibinfo{author}{\bibfnamefont{I.~K.} \bibnamefont{Schuller}},
  \bibinfo{journal}{Appl.\ Phys.\ Lett.} \textbf{\bibinfo{volume}{96}},
  \bibinfo{pages}{022513} (\bibinfo{year}{2010}).

\bibitem[{\citenamefont{Godfrey and Johnson}(2006)\citenamefont{Godfrey and Johnson}}]{Godfrey-Johnson2006}
\bibinfo{author}{\bibfnamefont{R.}~\bibnamefont{Godfrey}}
  \bibnamefont{and}
  \bibinfo{author}{\bibfnamefont{M.} \bibnamefont{Johnson}},
  \bibinfo{journal}{Phys.\ Rev.\ Lett.} \textbf{\bibinfo{volume}{96}},
  \bibinfo{pages}{136601} (\bibinfo{year}{2006}).

\bibitem[{\citenamefont{Kimura and Otani}(2007)\citenamefont{Kimura and Otani}}]{Kimura2007}
\bibinfo{author}{\bibfnamefont{T.}~\bibnamefont{Kimura}}
  \bibnamefont{and}
  \bibinfo{author}{\bibfnamefont{Y.} \bibnamefont{Otani}},
  \bibinfo{journal}{Phys.\ Rev.\ Lett.} \textbf{\bibinfo{volume}{99}},
  \bibinfo{pages}{196604} (\bibinfo{year}{2007}).

\bibitem[{\citenamefont{Casanova et~al.}(2009)\citenamefont{Casanova, Sharoni, Erekhinsky, Schuller}}]{Casanova2009}
\bibinfo{author}{\bibfnamefont{F.}~\bibnamefont{Casanova}},
\bibinfo{author}{\bibfnamefont{A.}~\bibnamefont{Sharoni}},
\bibinfo{author}{\bibfnamefont{M.}~\bibnamefont{Erekhinsky}},
\bibnamefont{and}
  \bibinfo{author}{\bibfnamefont{I.~K.} \bibnamefont{Schuller}},
  \bibinfo{journal}{Phys.\ Rev.\ B} \textbf{\bibinfo{volume}{79}},
  \bibinfo{pages}{184415} (\bibinfo{year}{2009}).

\bibitem[{\citenamefont{Mihajlovi\'{c} et~al.}()\citenamefont{Mihajlovi\'{c},
  Pearson, Garcia, Bader, and Hoffmann}}]{Mihajlovic2009}
\bibinfo{author}{\bibfnamefont{G.}~\bibnamefont{Mihajlovi\'{c}}},
  \bibinfo{author}{\bibfnamefont{J.~E.} \bibnamefont{Pearson}},
  \bibinfo{author}{\bibfnamefont{M.~A.} \bibnamefont{Garcia}},
  \bibinfo{author}{\bibfnamefont{S.~D.} \bibnamefont{Bader}}, \bibnamefont{and}
  \bibinfo{author}{\bibfnamefont{A.}~\bibnamefont{Hoffmann}},
  \bibinfo{journal}{Phys.\ Rev.\ Lett.}~\textbf{\bibinfo{volume}{103}},
  \bibinfo{pages}{166601} (\bibinfo{year}{2009}).

\bibitem[{\citenamefont{Jedema et~al.}(2002)\citenamefont{Jedema, Heersche, Filip, Baselmans, and van Wees}}]{Jedema2002}
\bibinfo{author}{\bibfnamefont{F.~J.}~\bibnamefont{Jedema}},
\bibinfo{author}{\bibfnamefont{H.~B.}~\bibnamefont{Heersche}},
\bibinfo{author}{\bibfnamefont{A.~T.}~\bibnamefont{Filip}},
\bibinfo{author}{\bibfnamefont{J.~J.~A.}~\bibnamefont{Baselmans}},
\bibnamefont{and}
  \bibinfo{author}{\bibfnamefont{B.~J.} \bibnamefont{van Wees}},
  \bibinfo{journal}{Nature\ (London)} \textbf{\bibinfo{volume}{416}},
  \bibinfo{pages}{713} (\bibinfo{year}{2002}).

\bibitem[{\citenamefont{Valenzuela and Tinkham}(2006)}]{Valenzuela2006}
\bibinfo{author}{\bibfnamefont{S.~O.} \bibnamefont{Valenzuela}}
  \bibnamefont{and} \bibinfo{author}{\bibfnamefont{M.}
  \bibnamefont{Tinkham}}, \bibinfo{journal}{Nature (London)}
  \textbf{\bibinfo{volume}{442}}, \bibinfo{pages}{176} (\bibinfo{year}{2006}).

\bibitem{Stoner1948}
\bibinfo{author}{\bibfnamefont{E.~C.}~\bibnamefont{Stoner}}
  \bibnamefont{and}
  \bibinfo{author}{\bibfnamefont{E.~P.} \bibnamefont{Wohlfarth}},
  \bibinfo{journal}{Philos.\ Trans.\ R.\ Soc.\ London, Ser.\ A} \textbf{\bibinfo{volume}{240}},
  \bibinfo{pages}{599} (\bibinfo{year}{1948}).

\bibitem[{\citenamefont{Mihajlovi\'{c} et~al.}(2010)\citenamefont{Mihajlovic, Dan Schreiber, Yuzi Liu, Amanda Petford-Long, Pearson, Garcia, Bader, and Hoffmann}}]{Mihajlovic2010}
\bibinfo{author}{\bibfnamefont{G.}~\bibnamefont{Mihajlovi\'{c}}},
\bibinfo{author}{\bibfnamefont{D.}~\bibnamefont{Schreiber}},
\bibinfo{author}{\bibfnamefont{Y.}~\bibnamefont{Liu}},
\bibinfo{author}{\bibfnamefont{J.~E.}~\bibnamefont{Pearson}},
\bibinfo{author}{\bibfnamefont{S.~D.}~\bibnamefont{Bader}},
\bibinfo{author}{\bibfnamefont{A.}~\bibnamefont{Petford-Long}},
\bibnamefont{and}
\bibinfo{author}{\bibfnamefont{A.} \bibnamefont{Hoffmann}},
  \bibinfo{note}{unpublished}.

\bibitem[{\citenamefont{Shang et~al.}(1998)\citenamefont{Shang, Nowak, Jansen, Moodera}}]{Shang1998}
\bibinfo{author}{\bibfnamefont{C.~H.}~\bibnamefont{Shang}},
\bibinfo{author}{\bibfnamefont{J.}~\bibnamefont{Nowak}},
\bibinfo{author}{\bibfnamefont{R.}~\bibnamefont{Jansen}},
\bibnamefont{and}
  \bibinfo{author}{\bibfnamefont{J.~S.} \bibnamefont{Moodera}},
  \bibinfo{journal}{Phys.\ Rev.\ B} \textbf{\bibinfo{volume}{58}},
  \bibinfo{pages}{R2917} (\bibinfo{year}{1998}).

\bibitem[{\citenamefont{Fabian and Das Sarma}(1999)\citenamefont{Fabian and
  Das Sarma}}]{Fabian-DasSarma1999}
\bibinfo{author}{\bibfnamefont{J.}~\bibnamefont{Fabian}}
  \bibnamefont{and}
  \bibinfo{author}{\bibfnamefont{S.} \bibnamefont{Das Sarma}},
  \bibinfo{journal}{J.\ Vac.\ Sci.\ Technol.\ B} \textbf{\bibinfo{volume}{17}},
  \bibinfo{pages}{1708} (\bibinfo{year}{1999}).

\bibitem[{\citenamefont{Lindelof and Wang}(1986)\citenamefont{Lindelof and Wang}}]{Lindelof1986}
\bibinfo{author}{\bibfnamefont{P.~E.}~\bibnamefont{Lindelof}}
  \bibnamefont{and}
  \bibinfo{author}{\bibfnamefont{S.} \bibnamefont{Wang}},
  \bibinfo{journal}{Phys.\ Rev.\ B} \textbf{\bibinfo{volume}{33}},
  \bibinfo{pages}{1478} (\bibinfo{year}{1986}).

\bibitem[{\citenamefont{Wang and Xiao}(1995)\citenamefont{Wang and Xiao}}]{Wang1995}
\bibinfo{author}{\bibfnamefont{J.~-Q.}~\bibnamefont{Wang}}
  \bibnamefont{and}
  \bibinfo{author}{\bibfnamefont{G.} \bibnamefont{Xiao}},
  \bibinfo{journal}{Phys.\ Rev.\ B} \textbf{\bibinfo{volume}{51}},
  \bibinfo{pages}{5863} (\bibinfo{year}{1995}).

\bibitem[{\citenamefont{Sondheimer}(1952)\citenamefont{Sonheimer}}]{Sondheimer1952}
\bibinfo{author}{\bibfnamefont{E.~H.}~\bibnamefont{Sondheimer}},
  \bibinfo{journal}{Adv.\ Phys.} \textbf{\bibinfo{volume}{1}},
  \bibinfo{pages}{1} (\bibinfo{year}{1952}).

\bibitem[{\citenamefont{Watts and Cousins}(1968)\citenamefont{Watts and Cousins}}]{Watts1968}
\bibinfo{author}{\bibfnamefont{A.~J.}~\bibnamefont{Watts}}
  \bibnamefont{and}
  \bibinfo{author}{\bibfnamefont{J.~E.} \bibnamefont{Cousins}},
  \bibinfo{journal}{Phys.\ Stat.\ Sol.} \textbf{\bibinfo{volume}{30}},
  \bibinfo{pages}{105} (\bibinfo{year}{1968}).

\bibitem[{\citenamefont{Wang and Schumacher}(1973)\citenamefont{Wang and Schumacher}}]{Wang1973}
\bibinfo{author}{\bibfnamefont{S.~-K.}~\bibnamefont{Wang}}
  \bibnamefont{and}
  \bibinfo{author}{\bibfnamefont{R.~T.} \bibnamefont{Schumacher}},
  \bibinfo{journal}{Phys.\ Rev.\ B} \textbf{\bibinfo{volume}{8}},
  \bibinfo{pages}{4119} (\bibinfo{year}{1973}).

\bibitem[{\citenamefont{Zamaleev and Kharakhashyan}(1978)\citenamefont{Zamaleev and Kharakhashyan}}]{Zamaleev1978}
\bibinfo{author}{\bibfnamefont{I.~G.}~\bibnamefont{Zamaleev}}
  \bibnamefont{and}
  \bibinfo{author}{\bibfnamefont{E.~G.} \bibnamefont{Kharakhashyan}},
  \bibinfo{journal}{JETP\ Lett.} \textbf{\bibinfo{volume}{27}},
  \bibinfo{pages}{641} (\bibinfo{year}{1978}).

\bibitem[{\citenamefont{Dingle}(1950)\citenamefont{Dingle}}]{Dingle1950}
\bibinfo{author}{\bibfnamefont{R.~B.}~\bibnamefont{Dingle}},
  \bibinfo{journal}{Proc.\ R.\ Soc.\ London, Ser.\ A} \textbf{\bibinfo{volume}{201}},
  \bibinfo{pages}{545} (\bibinfo{year}{1950}).

\bibitem[{\citenamefont{Beuneu and Monod}(1976)\citenamefont{Beuneu and Monod}}]{Beuneu1976}
\bibinfo{author}{\bibfnamefont{F.}~\bibnamefont{Beuneu}}
  \bibnamefont{and}
  \bibinfo{author}{\bibfnamefont{P.} \bibnamefont{Monod}},
  \bibinfo{journal}{Phys.\ Rev.\ B} \textbf{\bibinfo{volume}{13}},
  \bibinfo{pages}{3424} (\bibinfo{year}{1976}).

\end{thebibliography}

\end{document}